\documentclass[aip, amsmath,amssymb, 10pt, twocolumn, shortbibliography]{revtex4-1}

\usepackage{graphicx} 
\usepackage{amsmath}
\usepackage{twoopt}
\usepackage{multirow}
\usepackage{caption}
\usepackage{siunitx}
\usepackage{threeparttablex}
\usepackage{subfig}
\newcommand\unity{1\!\!1}

\begin{document}

\title{Quantum study of the CH$_3^+$  photodissociation in full dimension Neural Networks potential energy surfaces}

\author{Pablo del Mazo-Sevillano}
\affiliation{Unidad Asociada UAM-IFF-CSIC,
          Departamento de Qu{\'\i}mica F{\'\i}sica Aplicada, Facultad de
          Ciencias M-14, Universidad Aut{\'o}noma de Madrid, 28049, Madrid, Spain}
\author{Alfredo Aguado}
\affiliation{Unidad Asociada UAM-IFF-CSIC,
          Departamento de Qu{\'\i}mica F{\'\i}sica Aplicada, Facultad de
          Ciencias M-14, Universidad Aut{\'o}noma de Madrid, 28049, Madrid, Spain}
    
  \author{Javier R. Goicoechea} 
  \affiliation{Instituto de F{\'\i}sica Fundamental (IFF-CSIC), C.S.I.C.,                                   
         Serrano 123, 28006 Madrid, Spain}
  \author{Octavio Roncero}
  \email{octavio.roncero@csic.es}
  \affiliation{Instituto de F{\'\i}sica Fundamental (IFF-CSIC), C.S.I.C.,                                   
         Serrano 123, 28006 Madrid, Spain}

\date{\today}


     \begin{abstract}
       CH$_3^+$, a cornerstone intermediate in interstellar chemistry,  has recently been detected for the first time
       by the James Webb Space Telescope. The photodissociation of this ion is studied here.
       Accurate explicitly correlated multi-reference  configuration interaction {\it ab initio} calculations are done,
       and full dimensional potential energy surfaces are developed for the three lower electronic states,
       with a fundamental invariant neural network method. The photodissociation cross section is calculated
       using a full dimensional quantum wave packet method, in heliocentric Radau coordinates. 
       The wave packet is represented in angular and radial grids allowing to reduce the number of points physically accessible, requiring to push up the spurious states appearing when
       evaluating the angular kinetic terms, through a projection technique.
       The photodissociation spectra,
       when employed in astrochemical models to simulate the conditions of the Orion Bar,
       results in a lesser destruction of CH$_3^+$ compared to that obtained when utilizing the
       recommended values in the kinetic database for astrochemistry (KIDA).
     \end{abstract}
\maketitle

\section{Introduction}
 The long-sought-after CH$_3^+$ cation  has been recently detected for the first time in a
  protoplanetary disk (d203-506) 
  illuminated by the strong  far ultraviolet (FUV) radiation field from nearby massive stars in Orion's Trapezium cluster \cite{Berne-etal:23}. This detection was only possible
  in the infrared,  through vibrational spectroscopy, at $\approx$ 1400 cm$^{-1}$, within the PDRs4All program using the James Webb Space Telescope (JWST).
  This highly symmetric cation, with a planar $D_{3h}$ configuration, has no permanent dipole moment and thus cannot
  be observed through microwave rotational spectroscopy.
  On the contrary, the rotational spectra of its deuterated isotopologues, such as CH$_2$D$^+$ or CHD$_2^+$, has
  been experimentally characterized \cite{Gartnet-etal:13,Jusko-etal:17,Topfer-etal:18}, but only a tentative detection of  CH$_2$D$^+$
  has been reported so far \cite{Roueff-etal:13}.

  Hydrides are the first molecules to form in the interstellar medium (ISM) and provide crucial information
  on the physical conditions, such as the cosmic-ray ionization rate and H$_2$/H  abundance ratios \cite{Gerin-etal:16}.
  The precise determination
  of their abundances is key to the following chemistry in the ISM.
  Carbon hydrides are of paramount importance because the allotropy of carbon 
  triggers the molecular complexity in space: from organic and prebiotic molecules, to polycyclic aromatic hydrocarbons (PAH's),
  amorphous carbon  and many different minerals.
  CH$_n^+$ cations are particularly important because ion-molecule reactions are typically faster
  and the low ionization energy of carbon (11.3 eV), below that of hydrogen (13.6 eV), 
  produces   high  C$^+$/C abundance ratios in molecular gas irradiated by FUV (6 eV $<$ E $<$ 13.6 eV).
  Carbon cations present very anomalous properties, giving rise 
  to the development of the field of the
  carbocation chemistry\cite{Prakash-Schleyer:97,Olah-Prakash:04}, where the spectroscopic characterization
  of these species, pioneered by Takeshi Oka \cite{Crofton-etal:88,Jagod-etal:92,White-etal:99,Wang-etal:13},  plays an important role
  not only in astrochemistry but also in combustion chemistry.

  The smallest CH$^+$ carbocation is formed in C + H$_3^+$ or C$^+$ + H$_2$ reactions. The
  reaction C$^+$ + H$_2$ is endothermic by $\approx$ 0.5 eV \cite{Gerlich-etal:87}, but
  it becomes exothermic for vibrationally excited H$_2(v>1)$ \cite{Hierl-etal:97}. 
  It is known that enhanced abundances of FUV-pumped vibrationally excited H$_2$ significantly increase the reactivity of H$_2$ in FUV-irradiated molecular clouds\cite{Tielens-Hollenbach:85,Sternberg:1995,Agundez-etal:10}, so-called photodissociation regions (PDRs).
  Indeed, observations of PDRs reveal the presence of vibrationally excited H$_2$
  up to $v=12$  in several interstellar PDR's\cite{Kaplan-etal:17,Kaplan-etal:21}.
  The use of quantum state-to-state rate constants in chemical formation and excitation models applied to the formation
  of CH$^+$ describes very well the observed rotational emission lines detected in  PDRs \cite{Zanchet-etal:13,Faure-etal:17}.

  Once CH$^+$ is formed, the successive addition of hydrogen atoms occurs via reactive collisions with H$_2$, in exothermic
  or nearly thermoneutral reactions of the type H$_2$ + CH$_n^+$ $\rightarrow$ H +  CH$_{n+1}^+$.
  This sequence stops at CH$_3^+$, because the reaction H$_2$ + CH$_3^+$ is very slow and no CH$_4^+$ products are observed
  in several experiments\cite{Smith-etal:82,Asvany-etal:04,Asvany-etal:18}. The reaction to form the floppy
  methane cation \cite{Signorell-Merkt:99,Worner-etal:06}, CH$_4^+$, is endothermic and is not expected to be formed in this hydrogenation
  sequence. Instead, CH$_4^+$ is probably formed from neutral CH$_4$ by photoionization or electron impact,
  and this cation can react again with H$_2$ to form
  CH$_5^+$ \cite{Asvany-etal:04b}, a very floppy cation whose infrared spectra have been widely
  studied\cite{Schreiner-etal:93,White-etal:99,Asvany-etal:15}.

  The relative stability of CH$_3^+$ with H$_2$ makes  this cation play an important
  role in the formation of more complex molecules \cite{Dalgarno:85,Wakelam-etal:10,Chabot-etal:20}.
  The deuteration of CH$_3^+$ is relatively fast \cite{Smith-etal:82,Asvany-etal:04,Asvany-etal:18}
  and its deuterated isotopologues are considered to be determinant in the gas phase formation
  of complex deuterated molecules, whose observed abundance is several orders of magnitude higher
  than expected based on the cosmic D/H ratio.
  Moreover, since the rovibrational spectra of CH$_3^+$ can be observed by JWST, CH$_3^+$ is expected
  to be a useful diagnostic to determine the physical conditions of FUV-irradiated  environments
  (from clouds to protoplanetary disks\cite{Berne-etal:23,Henning-etal:24}).

  The vibrational spectroscopy of CH$_3^+$ has been the subject of
  many theoretical \cite{Kraemer-Spirko:91,Yu-Sears:02,Nyman-Yu:19,Changala-etal:23}
  and experimental \cite{Crofton-etal:85,Crofton-etal:88,Jagod-etal:92,Asvany-etal:18} studies.
  The photoionization of the neutral methyl radical has also been studied using
  several techniques \cite{Blush-etal:93,Wiedmann-etal:94,Liu-etal:01,Schulenburg-etal:06,Taatjes-etal:08,Loison:10,Cunha-de-Miranda-etal:10,Gans-etal:10,Changala-etal:23},
  which also gives information of the
  rovibrational structure of the CH$_3^+$ cation.

  The photodissociation cross sections of CH$^+$, CH$_2^+$ and CH$_4^+$ have
  been reported previously \cite{Heays-etal:17}.
  However, there is only one study on the photodissociation of CH$_3^+$ 
  carried out nearly 50 years ago \cite{Blint-etal:76}, in which vertical
  excitation  was considered from the planar $D_{3h}$ equilibrium geometry on the ground electronic state.
  It was concluded that the oscillator strength leading to dissociation from the ground electronic state is very low.
  It is worth mentioning that, in the kinetic data base for astrochemistry (KIDA),
  the recommended rate constant for the photodissociation of CH$_3^+$ under the local mean interstellar
  radiation field is $2 \cdot 10^{-9}$ s$^{-1}$ (see also Ref.~\cite{Tielens:10}).
  The value of $2 \cdot 10^{-9}$ s$^{-1}$ is rather high according to the previous study \cite{Blint-etal:76},
  and it is therefore important to determine the photodissociation rate of CH$_3^+$ more accurately.
  
  The objective of this work is to study the photodissociation cross-section of CH$_3^+$
  using quantum full dimension dynamics  to properly assess the destruction of this cation under different FUV radiation fields.
  The work is distributted as follows.
  In section II the {\it ab initio} calculations of the first electronic states of CH$_3^+$ are described. The neural network fitting
  of the first three electronic states are described in section III. The vibrational bound states of the ground electronic
  states are described in section IV. The transition dipole moments and their fit are described in section V. The calculation of
  the photodissociation cross section are described in section VI, and their use in astrochemical models in section VII. Finally,
  section VIII is devoted to extract some conclusions.

  \section{Electronic states}

\begin{figure}[t]
\begin{center}
    \includegraphics[width=0.95\linewidth]{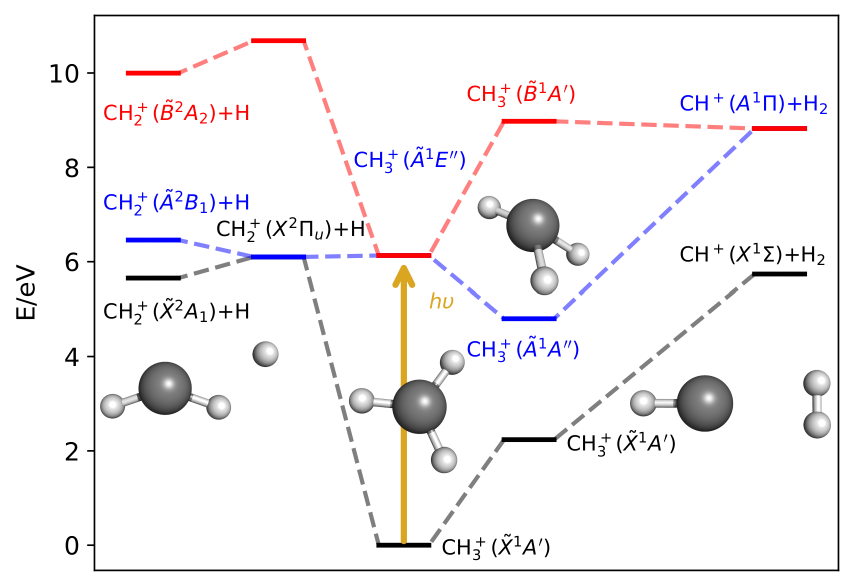}
\end{center}

    \caption{{Energy diagram of the lower electronic states of CH$_3^+$
        obtained with the ic-MRCI-F12 method. { The double degenerate CH$_2^+({ X}^2\Pi_u)$ +H,
          split in the bent ${\tilde X}$
        and ${\tilde A}$ states, produced by a
        strong Renner-Teller interaction.
 Black, blue and red lines refer to the ground, first and second excited electronic states,
    respectively}      }
  }
  \label{fig:mep}

\end{figure}

  The three lower electronic states of CH$_3^+$ are calculated using the explicitly correlated
  internally contracted multi-reference configuration interaction (ic-MRCI-F12)  method
   \cite{Werner-Knowles:88,Shiozaki-Werner:11}, with the MOLPRO suite of programs \cite{MOLPRO-WIREs} and the cc-pCVTZ-F12 electronic basis set\cite{Hill-etal:10}.
   The molecular orbitals are optimized using the state-averaged complete active space self-consistent field (SA-CASSCF) method, with 7 active orbitals,
   { for the three lower singlet electronic states}.
 Hereafter, the origin of energy is set at the planar $D_{3h}$ equilibrium configuration
 of the ground state, with a C--H distance of 1.0892 \AA,
 in very good agreement with
 previous calculations \cite{Blint-etal:76,Kraemer-Spirko:91,Yu-Sears:02,Delsaut:15}.
 An energy diagram of the three first electronic states is shown in Fig.~\ref{fig:mep}.
  
The ground electronic state correlates adiabatically with the CH$_2^+$(${\tilde X}^2A_1$) + H and CH$^+$($X^1\Sigma^+$) + H$_2$
asymptotes, which are both located at $\approx$ 6 eV over the equilibrium configuration. The ground and first excited
electronic states tend to the linear CH$_2^+$($X^2\Pi_u$) + H fragments, presenting a Renner-Teller interaction. 
The path towards the formation of CH$^+$ + H$_2$ can be seen as a subsequent step after the formation of CH$_2^+$ + H, where the H approaches one of the CH$_2^+$'s hydrogens, which is in an almost linear configuration. The C--H bond breaks while the H--H forms towards the CH$^+$+H$_2$ geometry. Due to the proximity of these geometries to the CH$_2^+$ linear configuration, the process occurs close to a conical intersection (CI).
The first adiabatic excited electronic states does not lead to CH$^+$ in the ground $X^1\Sigma^+$ state but in the excited $A^1\Pi$, a degenerate state
towards which the second excited state also correlates.

Considering a vertical excitation, the first electronic state corresponds to the double degenerate $^1E''$,
at the highly symmetric geometry of the ground equilibrium geometry,
as reported previously  \cite{Blint-etal:76,Delsaut:15}. The next
excited states in the Franck-Condon region, the $^1A_2''$ and $^1E'$, are $\approx$ 13 eV higher,
close to the atomic hydrogen ionization
and are not expected to contribute significantly. 

\begin{figure}[t]
\begin{center}
\includegraphics[width=0.95\linewidth]{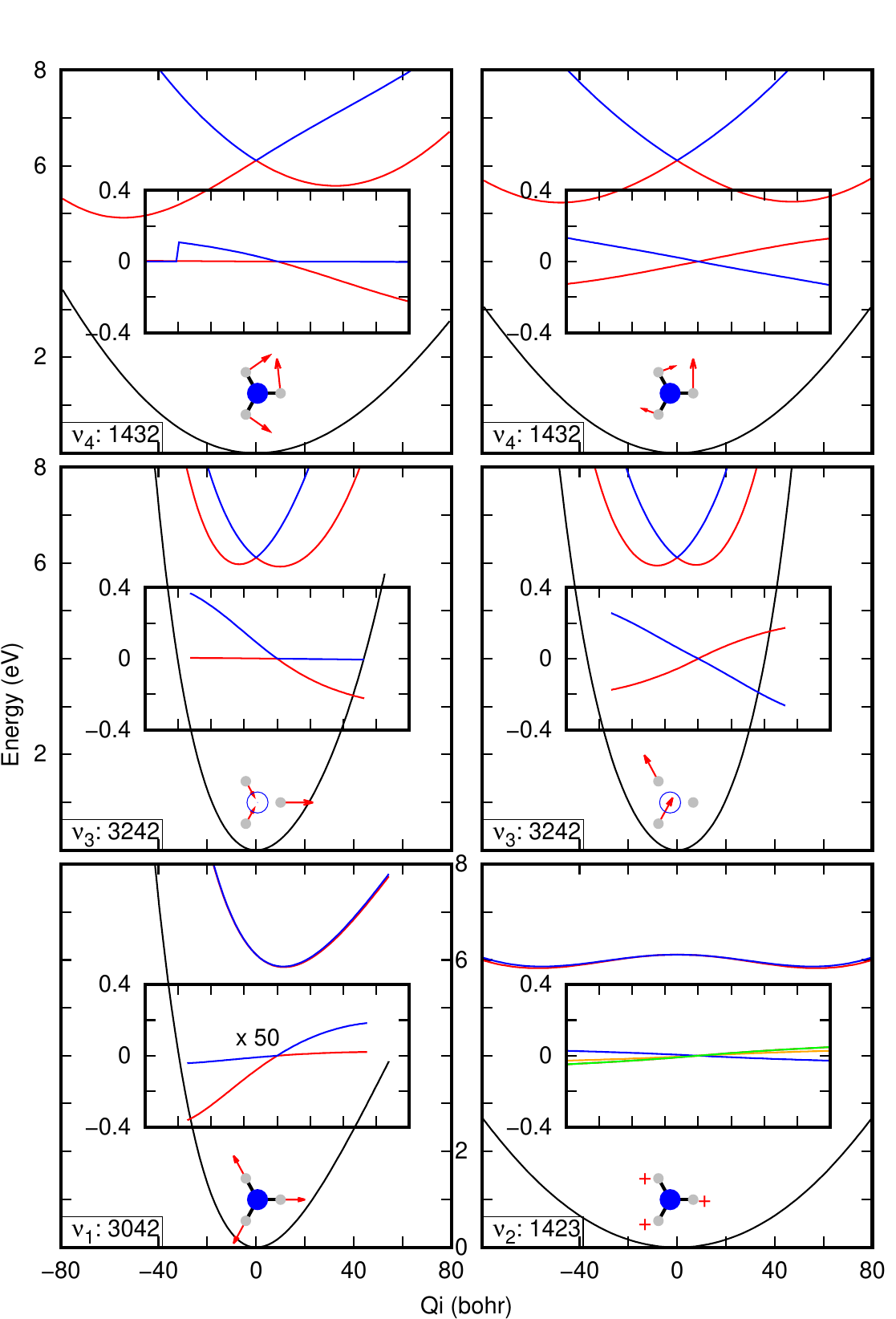}
\end{center}  
  \caption{Mono dimensional cuts along the normal modes, $Q_i$,
      of the CH$_3^+$ (at the planar D$_{3h}$ equilibrium geometry of the ground electronic state)
      for the lower three electronic states calculated at ic-MRCI-F12 level of theory.
      In each panel,
      the boxed inset corresponds to the transition electric dipole moment for the ${\tilde X}-{\tilde A}$ and ${\tilde X}-{\tilde B}$ transition (in atomic units)
      for the non-zero Cartesian components {(at equilibrium the molecule is in the
        x-y plane). For normal modes 2-6 only the z component is non-zero and red and blue lines
        correspond to the ${\tilde X}-{\tilde A}$ and
        ${\tilde X}-{\tilde B}$ transition dipole moment. For normal mode 1, the x, y components of $d_{XA}$
        and $d_{XB}$ are represented by red, orange, blue and green lines, respectively }. The other inset is a graphical representation of each normal mode.
      The energy of each normal mode is also indicated, in inverse centimeters.
}
 \label{fig:potNM} 
\end{figure}

The cuts of the potential along the normal coordinates of the ground state are shown in Fig.~\ref{fig:potNM}.
These normal modes are in good agreement with previously reported ones \cite{Blint-etal:76,Delsaut:15} and correspond to the singly
degenerate states, $\nu_1$, the symmetric stretching, and $\nu_2$, the umbrella vibration,
and two degenerate vibrations, $\nu_3$ and $\nu_4$. The elongation of the normal coordinates for $\nu_2$ and $\nu_1$ 
remains in the { $C_{3v}$} and { $D_{3h}$} symmetry, respectively, and the two excited electronic states remain degenerate.
This degeneracy is broken along the degenerate vibrations, $\nu_3$ and $\nu_4$, showing
the typical CI behavior { of the Jahn-Teller effect}, with the seam at the configuration of highest symmetry.

\section{Neural network potential energy sufaces fitting}

New analytical full dimensional potential energy surfaces (PESs) { have} been developed
to describe the three lower adiabatic electronic states of CH$_3^+$.
A fundamental invariant neural network (FI--NN)\cite{Shao2016} takes into account
the exact permutation symmetry of the three hydrogen atoms. Three FI--NN are trained --one for each of the three adiabatic energies. While a single FI--NN could handle the calculation of the three electronic states, this setup provides more flexibility in order to make use of the most accurate PESs for different tasks: vibrational state calculation in the ground electronic state and quantum dynamics in the excited states. Moreover, the data from the third excited state tends to be noisier due to interactions with higher excited states, what could interfere with the training process.

In all cases, the multilayer perceptron (MLP) architecture is used, which involves two hidden layers with 50 neurons each.
Hyperbolic tangents are used as activations between the hidden layers.
The input features are represented by fundamental invariants (FI) of the $p_{ij} = \exp(-\alpha \cdot d_{ij})$ functions, with $\alpha = 0.5$ $a_0^{-1}$ and $d_{ij}$ the interatomic distance between atoms $i$ and $j$. There is a total of nine FI for the A$_3$B case, which expressions are provided elsewhere~\cite{FIdef}.
The mathematical expression of the MLPs is the standard one, where the values of the $i$th neuron in the $(l+1)$ layer are computed through those from the previous layer and a trainable set of weights ($\boldsymbol{w}$) and bias ($\boldsymbol{b}$). $\sigma$ represents the activation function ---the hyperbolic tangent or linear function.
\begin{equation}
    H_i^{(l+1)} = \sigma \left(w_{ij}^{(l)} H_j^{(l)} + b_i^{(l)}\right)
\end{equation}

The MLPs are trained on a set of nearly 25000 energies computed
at a ic-MRCI-F12/cc-pCVTZ-F12 level of theory with MOLPRO 2012.
An extra set of about 5000 energies is left as test set.
A total of ten models are trained, but only the one which better performs
on the test set is used. Building this energy dataset is performed
in an iterative process, which starts with a rather small set of geometries
computed from normal mode displacements of equilibrium geometries and then
includes data from minimum energy paths or quantum and classical dynamics on
intermediate fits of the system.
These fits are done up to 15 eV and 25 eV
for the two first and third electronic states, respectively.

The training process is similar
to those previously described for H$_4^+$ and OH + H$_2$CO systems \cite{delMazoSevillano-etal:21,delMazo-Sevillano-etal:23}
and is performed with an in-house Python code based on
PyTorch library\cite{Paszke_PyTorch_An_Imperative_2019}.
An L-BFGS optimizer\cite{liuLimitedMemoryBFGS1989} is used
and the loss function is the Mean Square Error (MSE) error of the predicted energies
compared with the ic-MRCI-F12/cc-pCVTZ-F12 energies:
\begin{equation}
    \mathcal{L} = \frac{1}{N} \sum_{i=1}^{N} \left( E_i - E^*_i\right)^2
\end{equation}
where $N$ is the total amount of training data and $E_i$ is
the $i$th energy.
The asterisk indicates \textit{ab initio} energy.

The Root Mean Square Error (RMSE) of the three PES is presented in table~\ref{tab:rmse_pes} for several energy ranges. The errors are shown in meV units. The PES for the ${\tilde X}$ state remains accurate up to electronic energies of 6--7 eV, enough to compute highly accurate vibrational states. The PESs for the ${\tilde A}$ and ${\tilde B}$ states remain accurate up to higher electronic energies, although the latter presents larger errors, in part due to the difficulty to converge the {\it ab initio} calculations for this state, which interacts with higher excited electronic states.
\begin{table}[]
    \centering
    \begin{tabular}{l l l l}
    \hline
        E $<$ /eV & State ${\tilde X}$ & State ${\tilde A}$ & State ${\tilde B}$ \\
    \hline
        $1.0$ & $26.2$ $(461)$ & -- & --   \\
        $5.0$ & $46.5$ $(5960)$ & $43.8$ $(391)$ & --   \\
        $10.0$ & $131.3$ $(17657)$ & $92.7$ $(13428)$ & $135.5$ $(7528)$ \\
        $15.0$ & $355.8$ $(24998)$ & $91.4$ $(24803)$ & $150.2$ $(23230)$ \\
    \hline
    \end{tabular}
    \caption{RMSE for the PES of the three electronic states. The errors are presented in meV. In parenthesis the number of geometries in the energy range.}
    \label{tab:rmse_pes}
\end{table}

\begin{figure}
    \centering
    \includegraphics[width=0.95\linewidth]{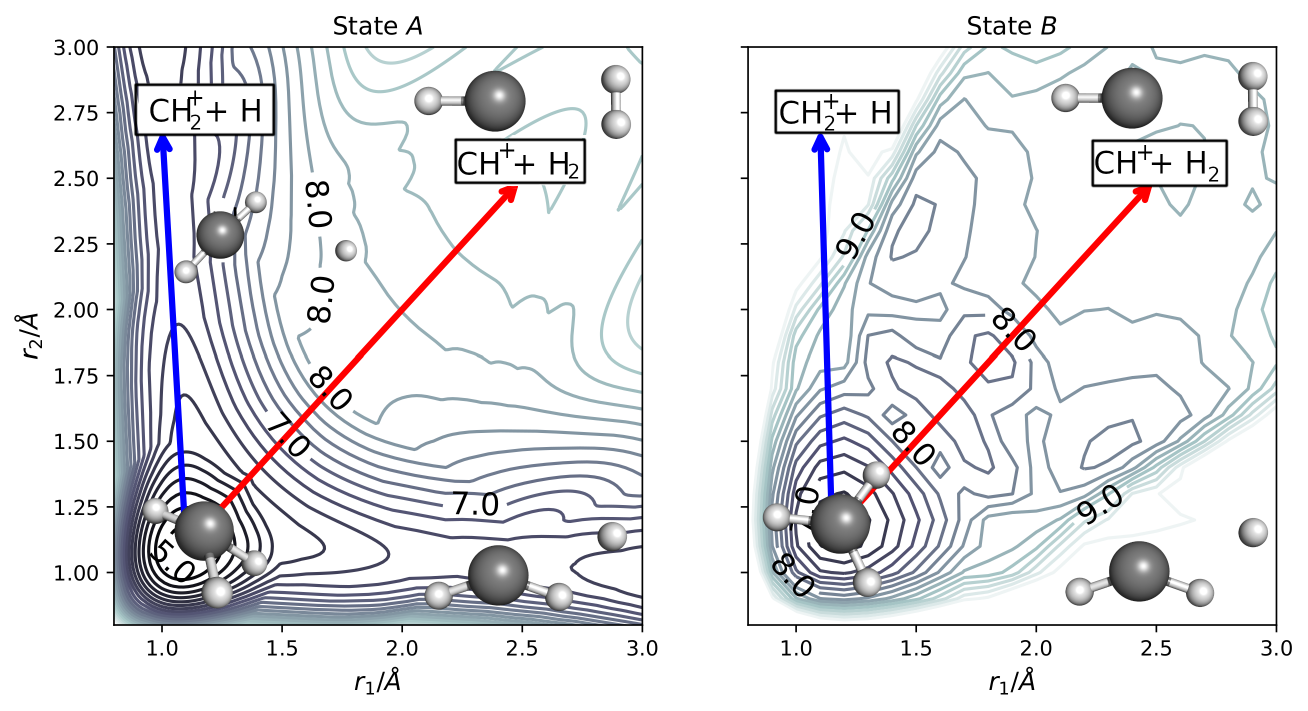}
    \caption{Contour plot of the PES for the ${\tilde A}$ and ${\tilde B}$ electronic states
      in terms of the heliocentric Radau coordinates $r_1$ and $r_2$.
      All the other coordinates are relaxed to the minimum energy configuration. The energies are represented in eV.}
    \label{fig:cuts}
\end{figure}

In the following we analyse in more detail the ${\tilde A}$ and ${\tilde B}$ states. Fig.~\ref{fig:cuts} presents the relaxed PES over two radial coordinates $r_1$ and $r_2$,
using heliocentric Radau coordinates as defined below. For the ${\tilde A}$ state there is a minimum, corresponding to a CH$_3^+$ ($2^1A$). The minimum in the ${\tilde B}$ state is in the Franck--Condon region and corresponds to CH$_3^+$ ($^1E''$).

The path to the CH$_2^+$ + H product occurs in the ${\tilde A}$ state after surpassing a low energy transition state less than 1 eV above the minimum. On the other hand, the path to the CH$^+$ + H$_2$ is highly endothermic, $\approx 4$ eV over the minimum, with no barrier. The path towards the formation of CH$_2^+$ + H can be merely explain as a C--H elongation ---related to the elongation of the $r_i$ coordinate in Fig~\ref{fig:cuts}. The path towards CH$^+$ + H$_2$ is not so direct, and proceeds via elongation of one of the C--H distances getting close to a CH$_2^+$ configuration. After this, the second $r_i$ distance elongates breaking a C--H bond while forming the H$_2$ molecule. In both cases the minimum energy paths get close to an almost linear configuration of the CH$_2^+$ ---a $^2\Pi_u$ state, degenerate with the ground electronic state--- what implies that ${\tilde X}$ and ${\tilde A}$ electronic states get close in energy as the photodissociation process occurs.

Regarding the reactions on the ${\tilde B}$ state we find that both reactions are highly endothermic. The Franck--Condon region becomes the absolute minimum with no other product close in energy as the CH$_2^+$ + H in the $A$ electronic state. Hence, we do not expect reactivity in this state to be important up to photon energies $\approx 9$ eV and $\approx 10$ eV for CH$^+$ + H$_2$ and CH$_2^+$ + H respectively. For this reason, we expect the CH$_3^+$ in the ${\tilde B}$ electronic state to remain mostly bounded for the photon energies of interest in this work.

\section{Bound vibrational states}

\begin{figure}[t]
\begin{center}
  \includegraphics[width=0.95\linewidth]{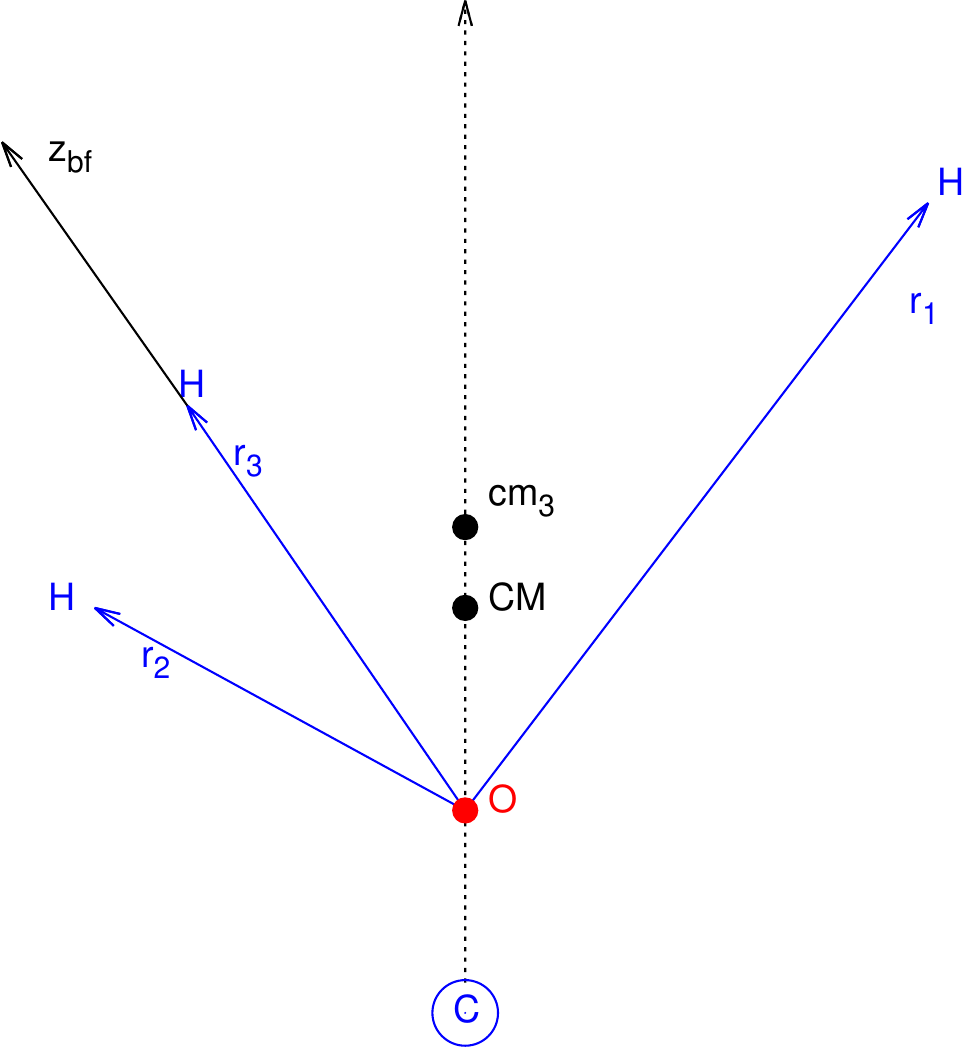}
  
  \caption{{Graphical description of the heliocentric Radau coordinates
      used in this work. CM and cm$_3$ are the center-of-mass of CH$_3^+$ and the H$_3$ subunit, respectively.
      The origin O is defined as in Ref.~\cite{Yu-Muckerman:02} to eliminate kinetic crossing terms in the
      kinetic energy operator.
   }}    
\label{fig:radau}
\end{center}
\end{figure}
The bound state calculations are done in two steps: first, the eigenvalues are calculated
using an iterative non-orthogonal Lanczos method\cite{Cullum-Willoughby:85}, and second, 
a conjugate gradient method\cite{Froberg,Wyatt:89} is used to obtain the eigenvectors.
These procedures are implemented in a parallel MPI form
using heliocentric Radau coordinates
\cite{Smith:80,Yu-Muckerman:02,Yu-Sears:02}, {illustrated in Fig.~\ref{fig:radau}.}
Three vectors ${\bf r}_i$ are defined, corresponding to the distance of each
hydrogen to a center O, situated along the line joining the centers-of-mass of CH$_3^+$ and H$_3$.
This center O is chosen to make zero the kinetic coupling terms among  the vectors ${\bf r}_i$,
and the Hamiltonian thus built is formally identical to that of Jacobi coordinates\cite{Smith:80,Yu-Muckerman:02}.
A body-fixed
frame is chosen, in which ${\bf r}_3$ lies parallel to the $z$-axis, and ${\bf r}_1$ is in the
$x-z$ body-fixed plane. Thus the coordinates are separated as three external Euler angles,
$\alpha,\beta,\gamma$, defining the body-fixed frame, and six internal coordinates $r_i$ $(i=1,2,3)$,
$\theta_j$ $(j=1,2)$ and $\phi$. The wave functions, for a given total angular momentum $J$, are  described as
\begin{eqnarray}
  \Psi^{JM} = \sqrt{ \frac{2J+1}{8\pi^2}} D^{J*}_{M\Omega}(\alpha,\beta,\gamma)
  \frac{\Phi^{JM}_\Omega(r_1,r_2,r_3,\theta_1,\theta_2,\phi)}{r_1 r_2 r_3},\nonumber\\
\end{eqnarray}
where $ D^{J*}_{M\Omega}$ are Wigner rotation matrices \cite{Zare-book}, with $M$ and $\Omega$ being
the projections of the total angular momentum ${\bf J}$ on the space-fixed and body-fixed frames
respectively.

The internal coordinates are described in grids.
Sinc Discrete Variable Representation (DVR) \cite{Colbert-Miller:92}
is used to describe the radial $r_i$ coordinates, non-orthogonal Gauss-Legendre DVR \cite{Corey-Lemoine:92,Corey-etal:93}
to describe $\theta_i$, and equispaced points in the interval $\left[0,2\pi\right]$ to describe $\phi$.
The evaluation of each kinetic term is done by transforming to  finite basis representation (FBR), where it
is analytical. 
This transformation is done sequentially, one internal coordinate by one,
to save computation time as it is done in other approaches representing the wave function
in the FBR and then transforming sequentially to the DVR to evaluate the potential\cite{Goldfield:00,Lin-Guo:02}.

Representing the wave functions in grids for internal coordinates has the advantage of saving many
points, the so called L-shaped grids\cite{Mowrey:91}, thus reducing considerably the memory and time requirements
of the calculations. However, the numerical sequential method done to transform from DVR to FBR, usually introduces
spurious states when evaluating the rotational kinetic terms  using
finite DVR grid points. To avoid this problem,  we
have developed a projection method to move the spurious states up, out of the physical energy interval of interest,
as described in the appendix.

The bound state calculations are done using a grid of 20 DVR points in the radial coordinates, $r_i$,
in the interval $\left[0.5,1.6793\right]$ \AA, 30 Gauss-Legendre points for $\theta_i$, and 61 points in $\phi$.
About 10$^4$ Lanczos iterations were done to converge the eigenvalues.

\begin{figure}[t]
\begin{center}
  \includegraphics[width=0.95\linewidth]{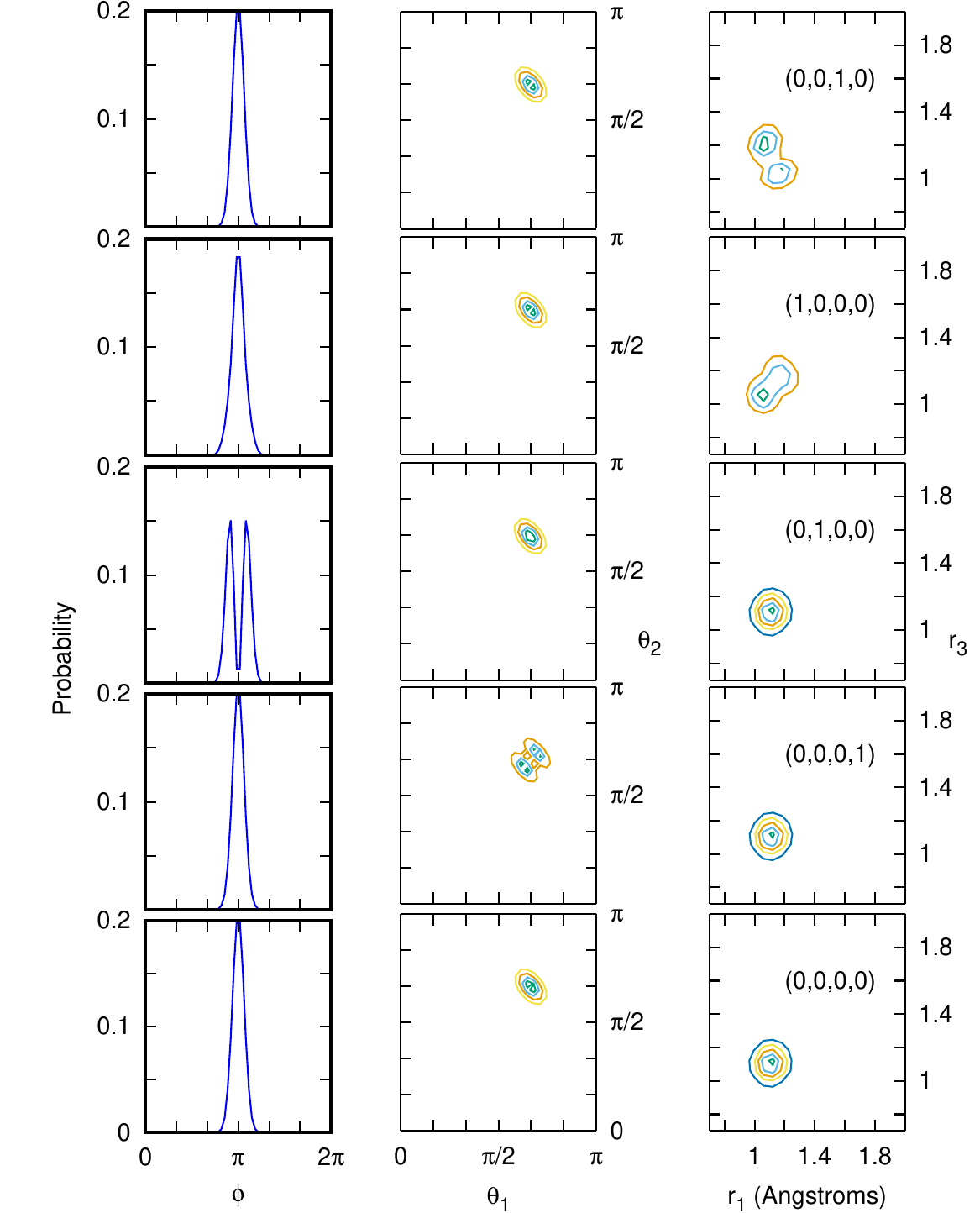}
  
  \caption{{Cuts of the density probabilities associated
  to different bound states of CH$_3^+({\tilde X})$, denoted by the 
  the vibrational modes  (${\nu_1}$, ${\nu_2}$, ${\nu_3}$, ${\nu_4}$). These bound
  states correspond to the the energy levels 1, 2, 3, 7 and 8.
   }}    
\label{fig:bound}
\end{center}
\end{figure}

Fig.~\ref{fig:bound} shows the cuts of the density probability associated to some bound states; those corresponding to the ground and first excitation on each mode (${\nu_1}$, ${\nu_2}$, ${\nu_3}$, ${\nu_4}$) for total angular momentum $J=0$. Their energies
are tabulated in Table.~\ref{tab:bound}. The heliocentric Radau coordinates are well adapted to describe the permutation symmetry of the hydrogen atoms, but in this first implementation no symmetry--adapted basis functions or grids are used. For
the degenerate modes (${\nu_3}$ and ${\nu_4}$) only one case is shown.

\begin{table}
  \caption{Lowest excitation energies (in cm$^{-1}$) for the vibrational modes of CH$_3^+$,
      obtained as described in the text for $J=0$. The zero-point energy of the ground rovibrational state is
    6776.898 cm$^{-1}$.}
\label{tab:bound}
\begin{tabular}{|c|c|c|c|}
    \hline
    vibrational mode  & This work & Ref.\cite{Kraemer-Spirko:91} & Ref.\cite{Changala-etal:23}  \\
    \hline
    $\nu_1$ & 2947.82 & 2949.8 & 2943.43 \\
    $\nu_2$ & 1424.53 & 1432.5 & 1405.72 \\
    $\nu_3$ & 3113.50 & 3091.3 & 3109.06 \\
    $\nu_4$ & 1393.91 & 1399.3 & 1394.98 \\
    \hline
  \end{tabular}
\end{table}

The lowest eigenvalues for each vibrational mode corresponding to
the bound states in the ground electronic state are listed in Table~\ref{tab:bound}, together with
other theoretical values for comparison. 
The values of Ref.~\cite{Changala-etal:23} were obtained to simulate the rovibrational
spectra observed in the d203-506 protoplanetary disk and experimental data.
The present results are within 5 cm$^{-1}$ accurate with respect to those previously
reported \cite{Kraemer-Spirko:91,Yu-Sears:02,Changala-etal:23}, except for the $\nu_2$ mode,
which deviates $\approx$ 20 cm$^{-1}$. Previous calculations are based on local fits for the potential,
thus describing only the bound states. In this work, however, the potential energy surfaces
are global, $i.e.$, they are built to describe the bound states and the fragmentation regions towards
the CH$^+$ + H$_2$ and the CH$_2^+$ + H products. For these reasons, we consider this new PES
as accurate enough to describe the photodissociation dynamics, with nearly spectroscopic accuracy.

\section{Transition dipole moments}

The transition dipole moments required for the ${\tilde X}-{\tilde A}$ and ${\tilde X}-{\tilde B}$ electronic excitation are calculated with MOLPRO
programs \cite{MOLPRO-WIREs}, and to avoid the randomness of the phase of adiabatic eigenvectors,
a biorthogonal transformation is used between consecutive points along lines. 
The Cartesian projections of the transition dipole moments are also shown,
in the boxed inset in  Fig.~\ref{fig:potNM}, for the ${\tilde X}-{\tilde A}$ and ${\tilde X}-{\tilde B}$ excitations,
 with the molecule being in the $x-z$ plane for the equilibrium geometry. In all cases, the transition dipole
 moments are zero at $Q_i=0$, corresponding to the equilibrium geometry.
 Only $\nu_2$ corresponds to a motion out of the
 plane of the planar $D_{3h}$ geometry,
 and it is the only one to have non zero components in $x$, $y$ and $z$ axis. For the rest
of the normal modes, only the component $y$, perpendicular to the plane of the molecule, is non-zero. This transition dipole
moment corresponds to a transition between two of the bonding orbitals of the C$^+$ atom (mostly corresponding to a $sp^2$ hybridization)
to an unoccupied $p_y$ orbital, out of the plane \cite{Delsaut:15}.
As a consequence of the CI of the ${\tilde A}$ and ${\tilde B}$ excited states
in $D_{3h}$ geometries\cite{Guan-etal:2020},
there is a sign change of the real electronic part of the wave function under a 2$\pi$ rotation
in the vibrational coordinates, a special case of Berry's geometrical phase.\cite{Herzberg-Longuet-Higgins:63,Berry:84,Changjian-et-al:2017}

The three components of both transition dipole moments have been fitted to an analytical function,
based on mono--dimensional grids for each internal heliocentric Radau coordinates. The fits are localized
in the CH$_3^+$ (${\tilde X}$) well, switching to zero outside this region.
There is no general method for an accurate representation of the dipole moment for polyatomic molecules,
using an appropriate functional form. Because the dipole moment is a vector property, its fit is more complicated than that for the energies\cite{Guan-etal:2020}. One alternative is to use a diabatic representation where the dipole moment is diagonal\cite{}.
In this work we are interested in fitting the adiabatic transition dipole moments between the ${\tilde X}^1A_1^\prime$ ground electronic state and the excited ${\tilde A}$ and ${\tilde B}$ ($^1E''$) states.

The phase of the adiabatic transition dipole moment $\mu_{ij}= \left\langle \phi_i | \hat{\mu} | \phi_j \right\rangle$ is arbitrary,
because it depends on the phase of the electronic wavefunctions $\phi_i$ and $\phi_j$. In addition, the adiabatic representation becomes inadequate near CIs \cite{Guan-etal:2020}, because real-valued adiabatic electronic wavefunction changes sign when transported around a CI (geometric or Berry phase)\cite{Berry:84,Changjian-et-al:2017}. In order to make the transition dipole moment continuous, we have
calculated the overlap of each electronic state with the same electronic state at a reference geometry --the equilibrium geometry of the ground $^1A'_1$ electronic state--
using the biorthogonalization method as programmed in MOLPRO program\cite{MOLPRO-WIREs}. The signs of
$\phi_i$ and $\phi_j$ are corrected in order to make the overlap positive, and then corrects the phase of $\mu_{ij}$. Therefore, in the adiabatic approximation we have not taken into account this change of sign of real electronic
wave functions that produces a change of sign of the transition dipole moment
when the conical intersection is surrounded in nuclear configuration space.

Once corrected the sign, in order to fit the transition dipole moment, we have expanded each component of the dipole moment as a function of symmetry coordinates of the $D_{3h}$ point group, defined in terms of the Heliocentric Radau coordinates defined above as
\begin{align*} \label{eq1}
S_1 & = \Delta r_1 + \Delta r_2 + \Delta r_3 \\
S_2 & = \Delta r_1 - \Delta r_2 \\
S_3 & = 2\Delta r_3 - \Delta r_1 - \Delta r_2 \\
S_4 & = \Delta \theta_1 + \Delta \theta_2 \\
S_5 & = \Delta \theta_1 - \Delta \theta_2 
\end{align*}
being $\Delta r_i=r_i-r_e \, (i=1,2,3)$ and $\Delta \theta_j=\theta_j-\theta_e \, (j=1,2)$ the variation with respect to equilibrium values, $r_e=1.089$~\AA~ and $\theta_e=2\pi/3$
and where $S_6$ is selected as the out-of-plane variation $\Delta \phi= \phi-\phi_e$ of the Radau angle with respect to the equilibrium value, $\phi_e=\pi$.

As shown in Fig.~\ref{fig:potNM}, the $\nu_1$ stretching mode corresponds to the variation of the $S_1$ symmetry coordinate, that belongs to the $A_1'$ irrep of the $D_{3h}$ point group. As a consequence, the only non-zero component is the out-of-plane $y$ component, although in this case it is practically zero, and can be discarded.
The $\nu_2$ bending mode corresponds to the variation of the out-of-plane coordinate. This mode belongs to the $A_2''$ irrep of the $D_{3h}$ point group. In this case the non-zero components are the $z$ component for the ${\tilde X}-{\tilde A}$ transition and the $x$ component for the ${\tilde X}-{\tilde B}$ transition. 
The other modes are degenerated, and corresponds to the $E'$ irrep. $\nu_3$ corresponds to stretching modes, which can be described by $S_2$ and $S_3$,
while $\nu_4$ correspond to bending modes that are described by $S_4$ and $S_5$. In this cases the non-zero component of the transition dipole is the $y$ component.

Since the $S_\alpha$ ($\alpha=2, ...,5$ ) coordinates do not take into account the symmetry properties of the dipole moment with respect to the exchange of two hydrogens, they are antisymmetrized as
$$
\widetilde{S}_\alpha= (-1)^s \sqrt[3]{|S_\alpha \cdot P_{13} S_\alpha \cdot P_{23} S_\alpha|}
$$
where $P_{ij}$ is the permutation operator for atoms $i$ and $j$ and where $s$ is a phase to take into account the symmetry of each component with respect to permutation of two hydrogens. When the dipole is antisymmetric with respect to the permutation, 
$(-1)^s$ is obtained as the sign of the maximum value of $S_\alpha, P_{13} S_\alpha$ or $P_{23} S_\alpha$.  
Finally, each Cartesian component of the transition dipole moment for the transition from ${\tilde X}^1A_1'$ to $({\tilde A},{\tilde B})^1E''$ states is expanded as a series in this symmetry coordinates $\widetilde{S}_\alpha$
$$
\mu_{ij}^{(x,y,z)} = \mu_{ij}^e + \sum_\alpha^6  a_{\alpha} \widetilde{S}_\alpha 
$$
with $\mu_{ij}^e=0$ in this case and where $a_\alpha$ are also developed as a serie
$$
a_\alpha = \left( \sum_{k}^{N_\alpha} a_{\alpha,k} \widetilde{S}_\alpha^k \right)
\cdot e^{-b_\alpha \widetilde{S}_\alpha^2}
$$
where $N_\alpha$ is the degree of the polynomial, and where the expansion has been multiplied by a Gaussian function in order to avoid extremely large values of the dipole moment in regions very far from the equilibrium position.

\begin{figure}[t]
\begin{center}
  \includegraphics[width=0.95\linewidth]{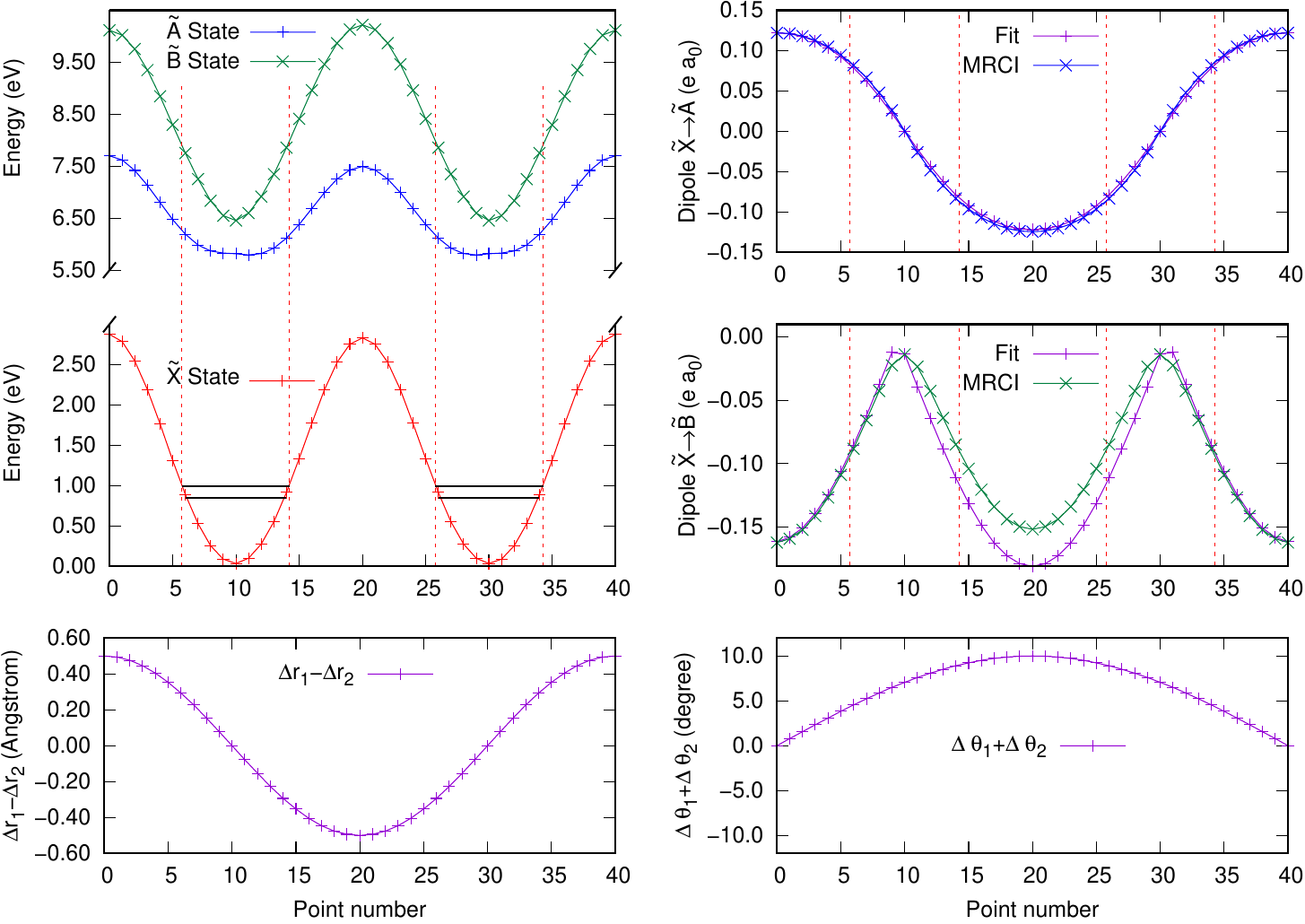}
  
  \caption{{Transition dipole moment $\mu^y$ as a function
      of the $S_2=\Delta r_1 -  \Delta r_2$ and $S_4=\Delta \theta_1 + \Delta\theta_2 $
      symmetry coordinates. { The vertical dashed lines defines the Frank-Condon region}
   }}    
\label{fig:dipole}
\end{center}
\end{figure}

In Fig.~\ref{fig:dipole} we show the variation of the transition dipole moment when the symmetry coordinates $S_2 = \Delta r_1 - \Delta r_2 $ and $S_4= \Delta \theta_1 + \Delta \theta_2$ are varied simultaneously, following a sinusoidal movement.

\section{Photodissociation cross section}

The photodissociation cross section is calculated for each transition with a wave packet method,
using the heliocentric Radau coordinate, as described above for the bound state calculations.
The modified Chebyshev propagator \cite{Mandelshtam-Taylor:95,Chen-Guo:96,Gray-Balint-Kurti:98,Gonzalez-Lezana-etal:05} is used
to integrate the Schr\"odinger equation as
\begin{eqnarray}\label{wvp-k}
\Phi(k=0)&=&\Psi(t=0) \nonumber \\
\Phi(k=1)&=& e^{-\varphi}{\hat H}_s \Phi(k=0)\\
\Phi(k+1)&=&e^{-\varphi} \left\lbrace 2 {\hat H}_s\Phi(k) -e^{-\varphi}\Phi(k-1)
             \right\rbrace,
             \nonumber
\end{eqnarray}
where ${\hat H}_s= \left( {\hat H} -E_0 \right)/ \Delta$ is the scaled Hamiltonian, with
 $E_0= (E_{max}+E_{min})/2$ and $\Delta E=
(E_{max}-E_{min})/2$, $E_{max}$ and $E_{min}$ being the minimum and
 maximum energy values of the Hamiltonian of the system represented
 in the grid/basis using in the propagation. 
 The wave packet at time $t$ and eigenfunctions at energy $E$ are expressed in terms of the Chebyshev
 iterations, $\Phi(k)$, as
 \begin{eqnarray}\label{wvf-chebyshev}
\Psi(t)&=& 
\sum_{k=0}^\infty f_k({\hat H}_s,t) \Phi(k) \nonumber \\
\Psi(E)&=& \sum_{k=0}^\infty c_k({\hat H}_s,E) \Phi(k)
\end{eqnarray}
with
\begin{eqnarray}
 f_k({\hat H}_s,t) &=& \left( 2-\delta_{k0}\right)  e^{-i E_0 t /\hbar}
 \,(-i)^k   \, J_k( t \Delta E /\hbar)\\
  c_k({\hat H}_s,E) &=& \left( 2-\delta_{k0}\right)
                        {\hbar \exp\left\lbrack-i k
                        \arccos\lbrace(E-E_0)/\Delta E\rbrace\right\rbrack\over \sqrt{\Delta E^2 -(E-E_0)^2}}
\nonumber                        
\end{eqnarray}
with $J_k$ being a Bessel function of the first kind. 

The absorption cross section is then given by
\begin{eqnarray}\label{eq:spectrum}
  \sigma( h\nu) &=& {A h\nu \over \pi\hbar} {\cal R}\int_0^\infty dt\,e^{-iEt/\hbar}
                    \,\left\langle   \Psi(t=0) \vert \Psi(t)\right\rangle  
 \\
  &=& {A h\nu \over \pi\hbar}  {\cal R} \sum_{k=0}^{\infty}  c_k({\hat H}_s,E)\,\left\langle   \Phi(k=0) \vert \Phi(k)\right\rangle 
      \nonumber
\end{eqnarray}
with $A=1/\hbar^2 \epsilon_0 c$ and ${\cal R}$ denoting the real part. For finite propagations (in this case 1000 Chebyshev iterations), the right--hand side of Eq.~\ref{eq:spectrum} is multiplied
by $\exp(- k \gamma)$, with $\gamma = 10^{-3}$ in the present case.

\begin{figure*}[t]
\begin{center}
  \includegraphics[width=0.45\linewidth]{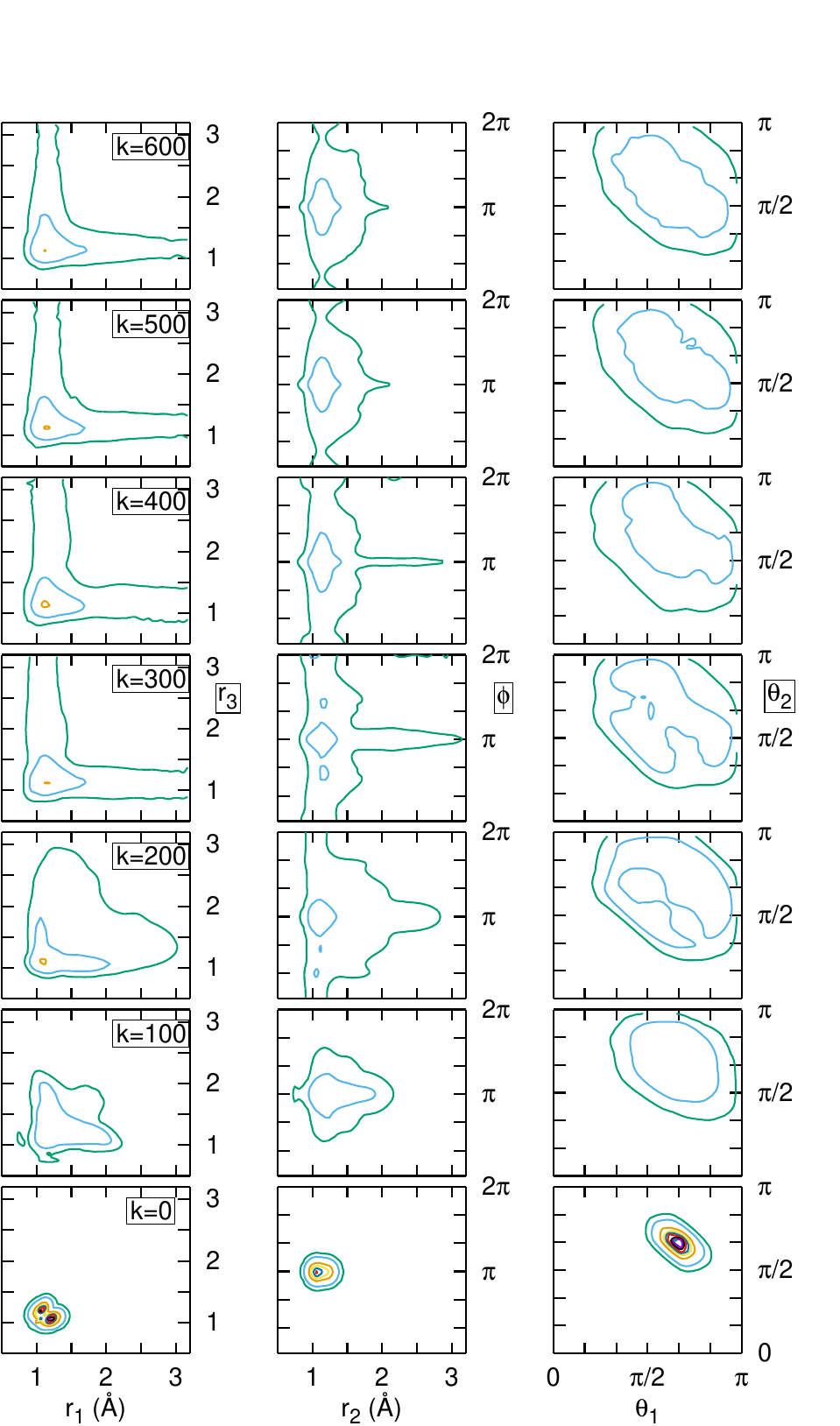}
  \includegraphics[width=0.45\linewidth]{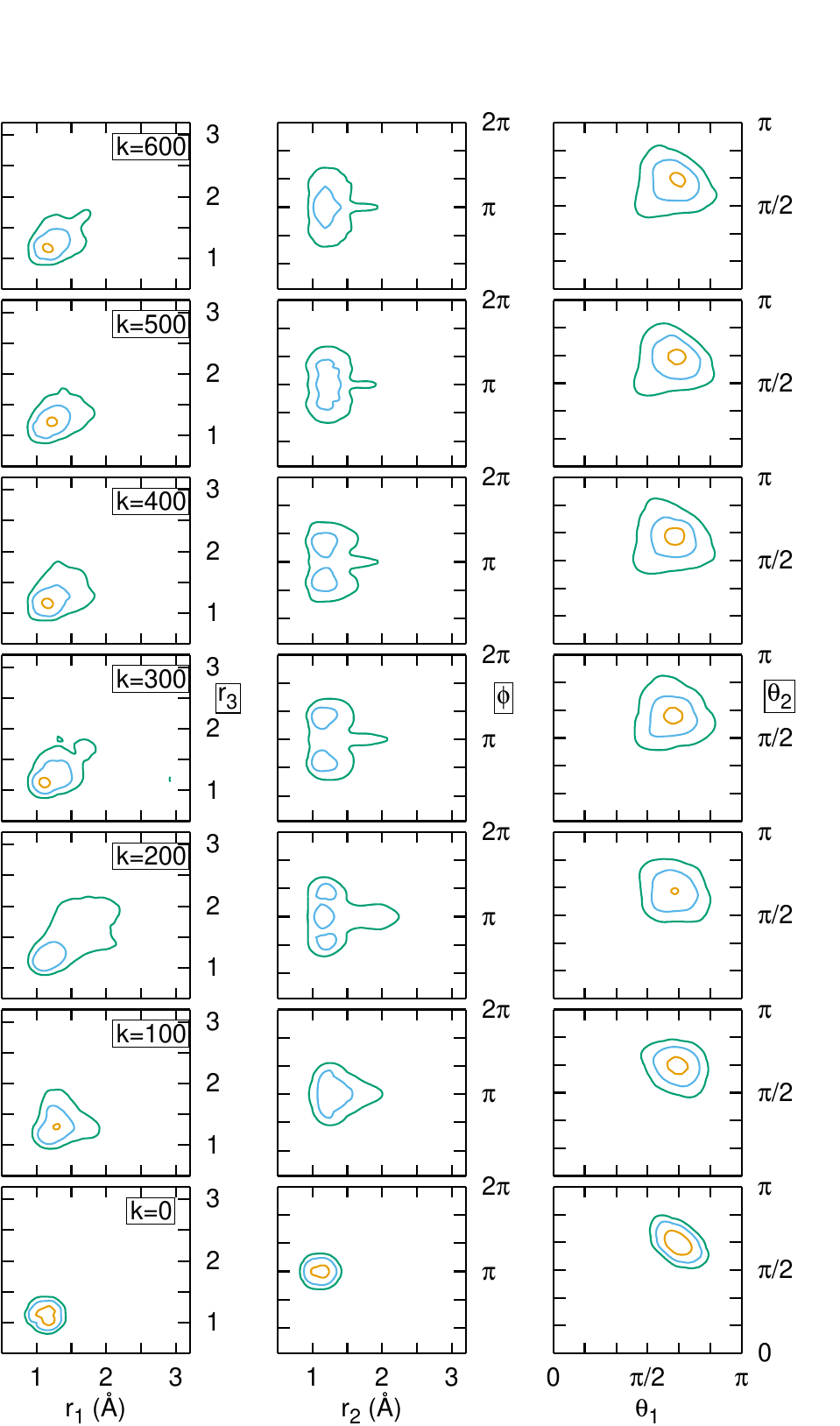}
  
  \caption{{Cuts of the density probability associated
  to the wave packet at different iterations $k$, for the ${\tilde A}$ (left panels) and ${\tilde B}$ (right panels) electronic states, for the transition from the ground electronic and vibrational state.
   }}    
\label{fig:wvp}
\end{center}
\end{figure*}

The wave packet is represented in grids for the internal radial and angular coordinates,
using the projection method to shift up the spurious solutions described in the appendix.
 The angular grids are those used for bound states,
while the radial grids are extended to 50 points, keeping the same density of points.
The initial wave packet is built for the $J_i$=0 $\rightarrow$ $J$=1 transition as described
previously\cite{Paniagua-etal:99,Aguado-etal:03,Chenel-etal:16}, combining the bound state
with the transition dipole moment to the final electronic states, { ${\tilde A}$ or ${\tilde B}$,} and projecting
on a final $J$. This is done for several bound states with different $\nu$ values.
The wave packet is propagated about
 1000 iterations.
At each iteration the autocorrelation function is evaluated, and photoabsorption
cross section is obtained by a Chebyshev transformation to the energy domain \cite{Gonzalez-Lezana-etal:05}.
     
In Fig.~\ref{fig:wvp}, contour plots of the density probability of the wave packet component, $\Phi(k)$,
are shown for the ${\tilde X}-{\tilde A}$ (left panels) and ${\tilde X}-{\tilde B}$ (right panels) transitions, for several values of $k$

\begin{figure}[t]
\begin{center}
  \includegraphics[width=0.95\linewidth]{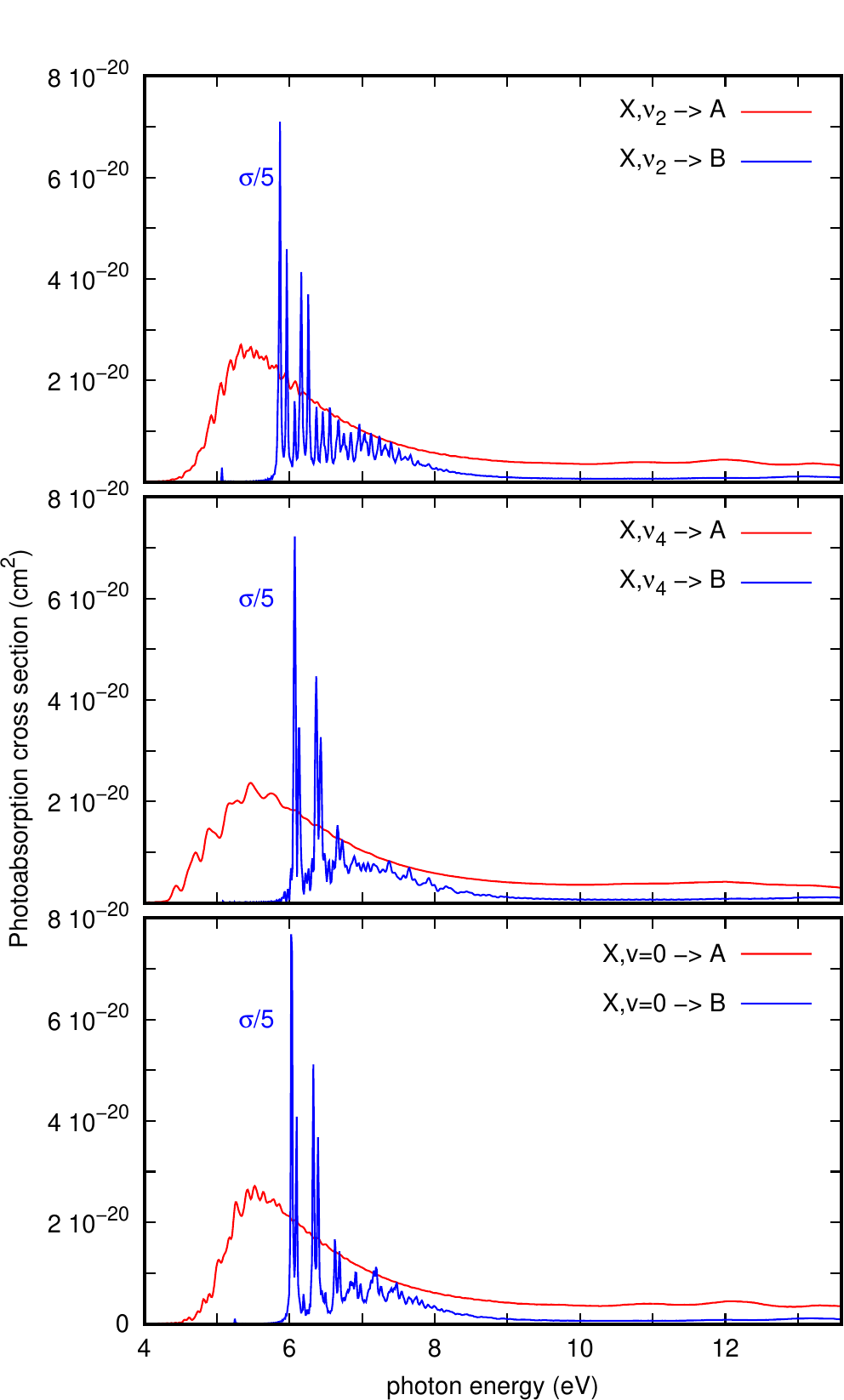}
  
  \caption{{CH$_3^+ $photoabsorption cross section (in cm$^2$)
      from the ground (bottom), $\nu_4$=1 (middle) and $\nu_2$=1 (top panel)  vibrational states towards
      the excited electronic states ${\tilde A}$ (red) and ${\tilde B}$ (blue), as a function of the
      photon energy (in eV). The cross section for the ${\tilde X}-{\tilde B}$ transitions 
      have been divided by 5 in the figure.
   }}    
\label{fig:espectra}
\end{center}
\end{figure}

The photoabsorption cross section towards the ${\tilde A}$ and ${\tilde B}$ electronic states are shown in Fig.~\ref{fig:espectra}
for different initial vibrational states. To explain the differences
between absorption to ${\tilde A}$ and ${\tilde B}$ states, it is important to remind that 
they present a conical intersection at the Franck-Condon region. Thus, the $A$
state corresponds to a local maximum which tends rapidly to the dissociation limits.
One of these limits corresponds to the CH$_2^+(X^2\Pi_u)$ + H products,
slightly below the vertical excitation. On the other side, towards the CH$^+(X^1\Sigma^+)$ + H$_2$ products there is an avoided crossing 
between the ${\tilde X}$ and ${\tilde A}$ state, from which the potential energy increases monotonically
towards the CH$^+(A^1\Pi)$ + H$_2$ asymptote, at 8.68 eV. The ${\tilde X}-{\tilde A}$ absorption spectrum shows a broad band
characteristic of a direct dissociation, mostly below photon energies of 8 eV
(corresponding to total energies of 8.84 eV). Clearly, the dissociation must be towards the lower CH$_2^+(X^2\Pi_u)$ + H products, which
is also supported by the inspection of the wave packet dynamics and the PESs. The ${\tilde X}-{\tilde A}$ absorption band shows some weak peaks at the lower energies
associated to resonances originated by the well around the minimum 
CH$_3^+({\tilde A}^1A_1')$ in Fig.~\ref{fig:mep}, which are above
the CH$_2^+(X^2\Pi_u)$ + H dissociation limit.

The upper part of the conical intersection, the ${\tilde B}$ state, corresponds
to a well, showing dissociation limits at 10.6 eV (CH$_2^+({\tilde B}^2 A_2)$ + H) and
8.68 eV (CH$^+(A^1\Pi)$ + H). Moreover, the PES shows a barrier of $\approx$ 10 eV when elongating one $r_i$ distance towards the
CH$_2^+(^2\Pi_u)$ + H products. As a consequence the ${\tilde X}-{\tilde B}$ absorption corresponds to resonant bound--bound transitions, with the wave packet
oscillating around the Franck-Condon regions showing many recurrences,
mostly at photon energies below 9 eV ($i.e.$ at total energies of $\approx$ 9.84 eV). Above this energy, the 
system can dissociate in the adiabatic ${\tilde B}$ state, what occurs with 
a low probability. Therefore, most of the wave packet should dissociate
by tunnelling at the CI, which tends mainly towards the CH$_2^+(^2\Pi_u)$ + H products. 

The spectra of vibrationally excited CH$_3^+$(${\tilde X}$, $\nu_i$=4,2)
shows very similar patterns. The ${\tilde X}-{\tilde A}$ bands for all vibrational states considered are very close, with a shift
in the photon energy of $\approx$ 1400 cm$^{-1}$ (0.174 eV)
between the ground and the two excited vibrational states.
The ${\tilde X}-{\tilde B}$ spectrum for $\nu_4$ shows a different intensity pattern 
as compared to that of the ground vibrational, as a result of the excitation on the $\theta_i$ angles. However, the ${\tilde X}-{\tilde B}$ for $\nu_2$ 
gets closer to that of the ground, what is explained by the shallower dependence of the potential on the out-of-plane angle $\phi$. 

\section{Astrochemical modeling}

The photodestruction of CH$_3^+$ in strongly FUV-irradiated objects
(such as interstellar PDRs and protoplanetary disks) is determined by the photodissociation rate, $i.e.$,
the integral of the photodissociation cross section with  energy dependent FUV radiation field.
Using Draine's \cite{Draine:78} mean interstellar radiation field, 
the CH$_3^+$ photodissociation rate  is $6.83 \cdot 10^{-12}$ s$^{-1}$, 7.24 $\cdot 10^{-12}$ s$^{-1}$
and 7.13 $\cdot 10^{-12}$ s$^{-1}$ for the ground vibrational state ($\nu=0$ in Fig.~\ref{fig:espectra}),
and for the $\nu_4=1$ and $\nu_2=1$ excited vibrational states, respectively,
with a very minor increase with vibrational excitation of $\approx$ 5-7 $\%$. These values are about 300 times
lower than the value of 
 $2\cdot 10^{-9}$ s$^{-1}$ recommended in KIDA data base.  Moreover,  KIDA 
 suggests that two photodestruction products, CH$_2^+$ and CH$^+$, form at the same rate,
 while according to this work the only product is CH$_2^+$ + H.

 The  photodissociation rate calculated here is rather low, in agreement
 with the previous estimation by Blint  and co-workers \cite{Blint-etal:76}.
The values reported for CH$^+$, CH$_2^+$ and CH$_4^+$ are 
$3.3 \cdot 10^{-10}$,  $1.4 \cdot 10^{-10}$ and 2$.8 \cdot 10^{-10}$ s$^{-1}$, respectively\cite{Heays-etal:17}. These rates are higher
than those obtained here for CH$_3^+$ by a factor of $\approx$ 30.
The reason for this is attributed to the ``forbidden'' nature of the transition dipole moment of CH$_3^+$ at the equilibrium configuration.

In interstellar clouds strongly illuminated by FUV photons, photoionization of carbon atoms
produces a high abundance of electrons, which rapidly recombine with cations,
producing excited neutral systems that dissociate. This dissociative recombination (DR) process is very fast, because 
of the  strong Coulomb interactions, of the order of 10$^{-7}$ s$^{-1}$.
Because of the large difference between the photodissociation and DR rates (of about 4 orders of magnitude),
it is expected that the destruction of CH$_3^+$ is dominated by electrons and not by photons.

\begin{figure}[t]
\begin{center}
  \includegraphics[width=0.95\linewidth]{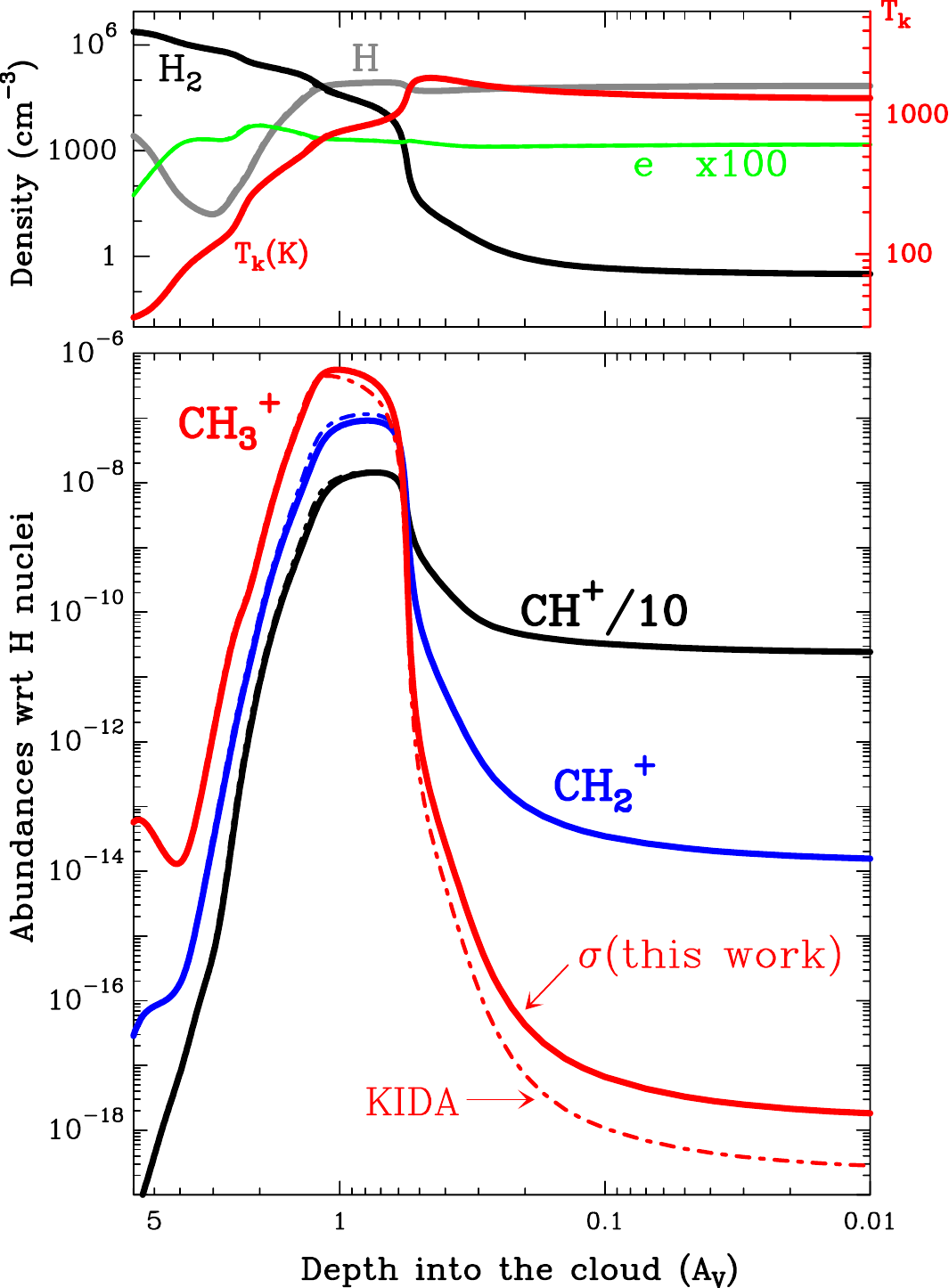}
\end{center}
  
  \caption{{ Results obtained with the Meudon PDR model using physical conditions
      corresponding to the Orion bar, as a function of FUV shielding
      or visual extinction parameter A$_V$ (low A$_V$  corresponds
      to the irradiated rim of the molecular cloud,
      while high A$_V$ correspond to distances well inside the molecular cloud
      with low FUV photon flux). Lower panel: abundance ratio
      (with respect to H) of CH$_n^+$ fractional abundances ($n$=1, 2 and 3),
      using the present CH$_3^+$ photodissociation rate (solid line)
      and that of KIDA data base (dashed lines).
      In the present case, the sum of ${\tilde X}-{\tilde A}$ and ${\tilde X}-{\tilde B}$
      photodissociation absorption yields to CH$_2^+$ products, as described in the text. Upper panel:
    Evolution of temperature and densities of H, H$_2$ and electrons and gas temperature as
  a function of A$_V$.
   }}    
\label{fig:abundance-Orion-bar}
\end{figure}

To show the effect of the photodissociation cross section obtained in this work,
and the competition with other processes,
Fig.~\ref{fig:abundance-Orion-bar}
shows an example obtained with the Meudon PDR model \cite{LePetit-etal:06,Goicoechea:07}
of a strongly FUV-irradiated molecular cloud, with a FUV radiation field $4 \times 10^4$ times
the mean interstellar radiation field in the solar neighbourhood,
and a constant thermal pressure $P / k_B = n \cdot T = 10^8$ K cm$^{-3}$.
These parameters are appropriate to the Orion Bar PDR, an irradiated rim of the Orion molecular cloud\cite{Goicoechea-etal:16}.
The upper panel of Fig.~\ref{fig:abundance-Orion-bar} shows the predicted gas density,
electron density, and temperature profile as a function of depth into the molecular cloud
(in magnitudes of visual extinction, A$_V$). The lower panel shows the resulting abundance profiles,
with respect to H nuclei, for the main species discussed in the text. The continuous curves refer to a model
that integrates the wavelength-dependent CH$_3^+$ photodissociation cross-sections determined in this work for the A and B electronic states and leading to
CH$_2^+$ as products. The dashed curve shows a model that uses the CH$_3^+$ photodissociation rate recommended in KIDA.
The dominant process destroying CH$_3^+$ is dissociative recombination with electrons,
thus the two models predict relatively similar abundance profiles. The role of CH$_3^+$ photodissociation
is more clearly seen at the very edge of the PDR, at low A$_V$, where the flux and energy of FUV photons is high.
Here, the model using the recommended rate in KIDA is not realistic and underestimates the CH$_3^+$ abundance by a factor of about 6. 
Such difference explains the need of realistic  evaluations of the rate constants used
in the astrochemical models.

\section{Conclusions}

A quantum treatment is developed to study the photodissociation
of the CH$_3^+$ cation below 13.6 eV. Accurate full dimension
PESs are generated using a FI-NN method for the three lower electronic
states based on ic-MRCI-F12/cc-pCVTZ-F12 {\it ab initio}.
The transition dipole moments are also fit locally in the region around
the equilibrium configuration covering the vibrational bound states
in the ground electronic state.

The bound states and wave packet dynamics are studied using heliocentric
Radau coordinates, well adapted to account for the permutation symmetry
of the three hydrogen atoms. A full grid representation of the internal
(radial and angular) coordinates is implemented,
allowing saving of memory and computation time due to
the L-shape method that allow to discard the grid points
with high energy out of the energy range of physical interest.
To do so, it was found necessary to apply a projection method
to push up the spurious states appearing when evaluating
the angular kinetic terms using a sequential transformation
from a non-direct DVR basis set to the FBR representation.
This is implemented in the home made MadWave4 code,
a MPI parallel code written in Fortran.

The calculated bound eigenvalues in the ground electronic states
are in good agreement with previous  theoretical and experimental ones.
The photodissociation cross section from several initial
vibrational states towards the excited ${\tilde A}$ and ${\tilde B}$ electronic states have been calculated
using a quantum wave packet method.
The initial vibrational excitation has little influence 
in the photodissociation dynamics and the calculated photodissociation rate is about 300 times lower than the recommended one in the KIDA data base for astrochemistry.

{ The possible fragmentation products  in the adiabatic representation is mostly
  towards the CH$_2^+$+ H products for the ${\tilde A}$ state. On the ${\tilde B}$ electronic state, however,
  most of the absorption spectrum corresponds to the bound region, and without including
  non-adiabatic transitions the wave packet cannot yield to dissociation. It is considered
  that this bound wave packet could be trasferred to the ${\tilde A}$ state, where it can dissociate.
  A diabatization of the electronic Hamiltonian is being done
  to consider the non-adiabatic transitions needed to a proper description of the branching ratios.
This is left for a future work}

The effect of the calculated cross section in interstellar regions strongly
illuminated by FUV photons is analyzed using the Meudon PDR code
applied to the Orion Bar as a prototype.
It is found that the dominant destruction mechanism of CH$_3^+$
is the dissociative recombination with electrons,
and that the use of the KIDA photodissociation rate
{ underestimates}  the CH$_3^+$ abundance,
demonstrating the need of realistic evaluation of rate constants in
astrochemical models.

\section{Supplementary Material}

The three Neural Network PESs, in fortran programs,
and the photodissociation cross section obtained for the ground vibrational state
obtained in this work are supplied in the Supplementary information, giving detailed information about
how to be used.

\section{Acknowledgements}

 The research leading to these results has received funding from
 MICIN (Spain) under grants PID2021-122549NB-C21,
 PID2021-122549NB-C22 and PID2019-106110GB-I00.
 We thank the PDRs4All-JWST-ERS team for their work in planning,
 obtaining, and calibrating the spectroscopic observations
 that led to the detection of interstellar CH$_3^+$. 

 \section{Data availability Statement}

The data that support the findings of this study are available from the corresponding author
upon reasonable request.

\appendix
\section{Projecting up spurious solutions}

We describe here a method to eliminate spurious states that
appear when using fdiscrete variable representation (DVR) and a sequential transformation
to the  finite basis representation (FBR)
to evaluate the angular kinetic terms.

\begin{figure}[t]
\begin{center}
  \includegraphics[width=0.9\linewidth]{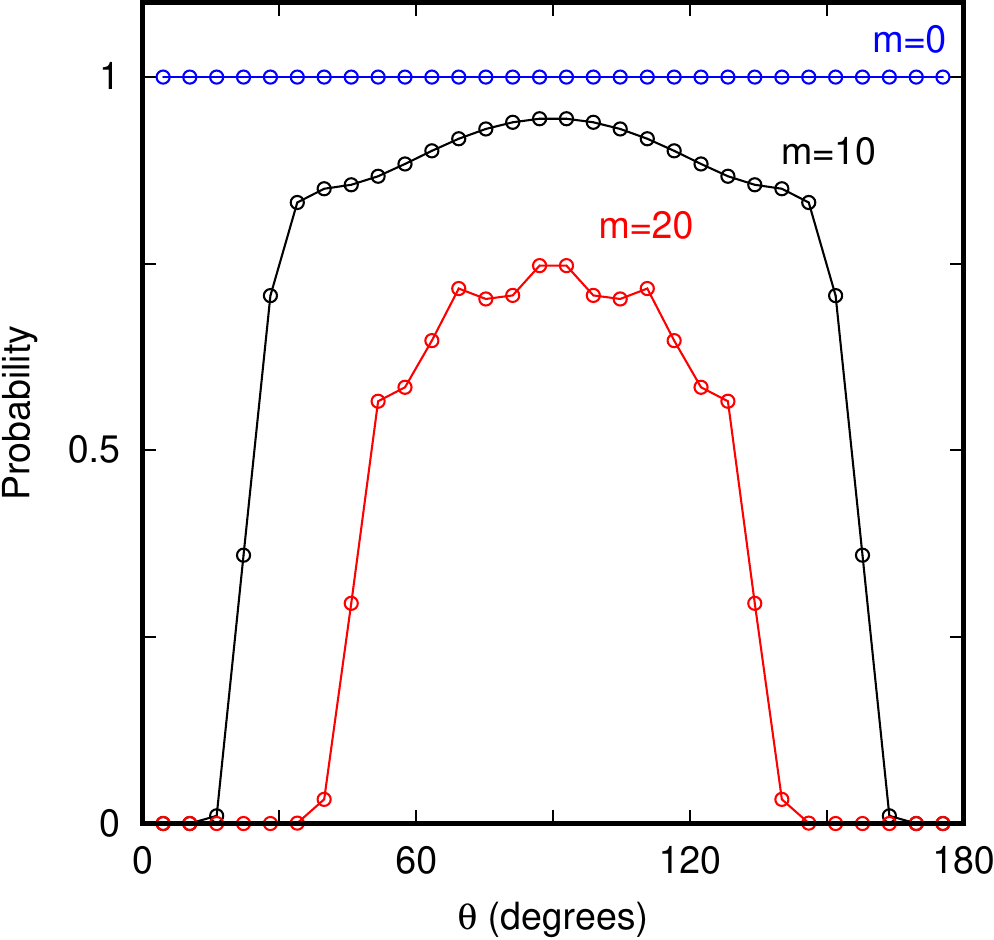}
  
  \caption{{ Closure represented in
      a grid as $ \langle \theta_i \vert  {\unity}^m \vert \theta_i\rangle $, for
      $m$=0, 10 and 20, for $j_{max}$=29 and a Gauss-Legendre grid of 30 points.
   }}    
\label{fig:m-closure}
\end{center}
\end{figure}

Spherical harmonics, $\vert j, m\rangle$, form a complete FBR set, and
are non-direct products of
functions in $\theta$ (normalized associated Legendre polynomials depending on the $m$ projection)
and $\phi$. 
The transformation to a  DVR in
$\theta$ and $\phi$ coordinates \cite{Corey-Lemoine:92},
formed by direct products of Gauss-Legendre points in $\theta$ and equispaced points in $\phi$,
is usually done in consecutive steps to reduce computational effort as \cite{Goldfield:00,Lin-Guo:02}
\begin{eqnarray}
\langle j, m   \vert \Psi \rangle \leftrightarrow \langle\theta_i, m  \vert  \Psi\rangle \leftrightarrow  \langle \theta_i, \phi_k \vert  \Psi \rangle,
\end{eqnarray}
where $\phi_k$ are equispaced points in the $\left[0,2\pi\right]$ and $\theta_i$ are Gauss-Legendre points in the 
$\left[0,\pi\right]$ interval, used for all projections $m$. In the intermediate
$\vert \theta_i, m \rangle$ representation $m$-independent Gauss-Legendre grid of points is not complete
for all $\theta_i$ values because at the extreme values the associated Legendre polynomials tends to zero
as $\sin^m\theta$. We can define a $m$-dependent closure relationship in a finite FBR and DVR representation as
\begin{eqnarray}
  \langle \theta_i \vert  {\unity}^m \vert \theta_i\rangle =
  \sum_{j=m}^{j_{max}} \langle \theta_i \vert j, m \rangle \langle j, m\vert\theta_i \rangle,
\end{eqnarray}
and a graphical representation  is shown in Fig.~\ref{fig:m-closure}.

Clearly, for $\theta_k$ near 0 and $\pi$ the closure relation is far from unity as $m$ increases, and this
introduces some spurious states using finite grids/basis.
When using the DVR-FBR transformation to evaluate rotational kinetic energy,
these spurious states will tend to have zero energy  and 
look like spikes. To remove these states in the physical window of the bound state or wave packet propagation,
these states are shifted up in energy by using the projector ${\cal P}_m={\unity}^0-{\unity}^m$.
To do so, once the wave function is expressed in the intermediate representation as $\langle  \theta_i,m \vert \Psi\rangle $,
the action of the rotational operator ${\bf j}^2$ takes the form
\begin{eqnarray}
&& \sum_{i'}  \langle  \theta_i, m \vert {\bf j}^2\vert \theta_{i'}, m \rangle \langle  \theta_{i'},m \vert \Psi\rangle  =\\
 &=&  \sum_{j=m} \langle  \theta_i, m \vert j,m \rangle j(j+1)\sum_{i'}  \,
 \langle j,m \vert \theta_{i'}m \rangle   \langle\theta_{i'},m \vert \Psi\rangle \nonumber\\
  &+& \sum_{j=0}  \,\langle  \theta_i, 0 \vert j,0 \rangle C_{max} \,\sum_{i'} 
  \langle j,0   \vert \theta_{i'},0 \rangle\langle\theta_{i'},m \vert \Psi\rangle\nonumber\\
  &-&\sum_{j=m}  \,\langle  \theta_i, m \vert j,m \rangle  C_{max} \,\sum_{i'} 
  \langle j,m   \vert \theta_{i'},m \rangle\langle\theta_{i'},m \vert \Psi\rangle\nonumber
 \end{eqnarray}
 where $C_{max}$ is a high positive constant, and here is chosen as the highest value of the potential energy.
 The three terms in the previous equation are evaluated as successive multiplication of a matrix and a vector,
 to save computation time.
 
\begin{figure}[t]
\begin{center}
  \includegraphics[width=0.9\linewidth]{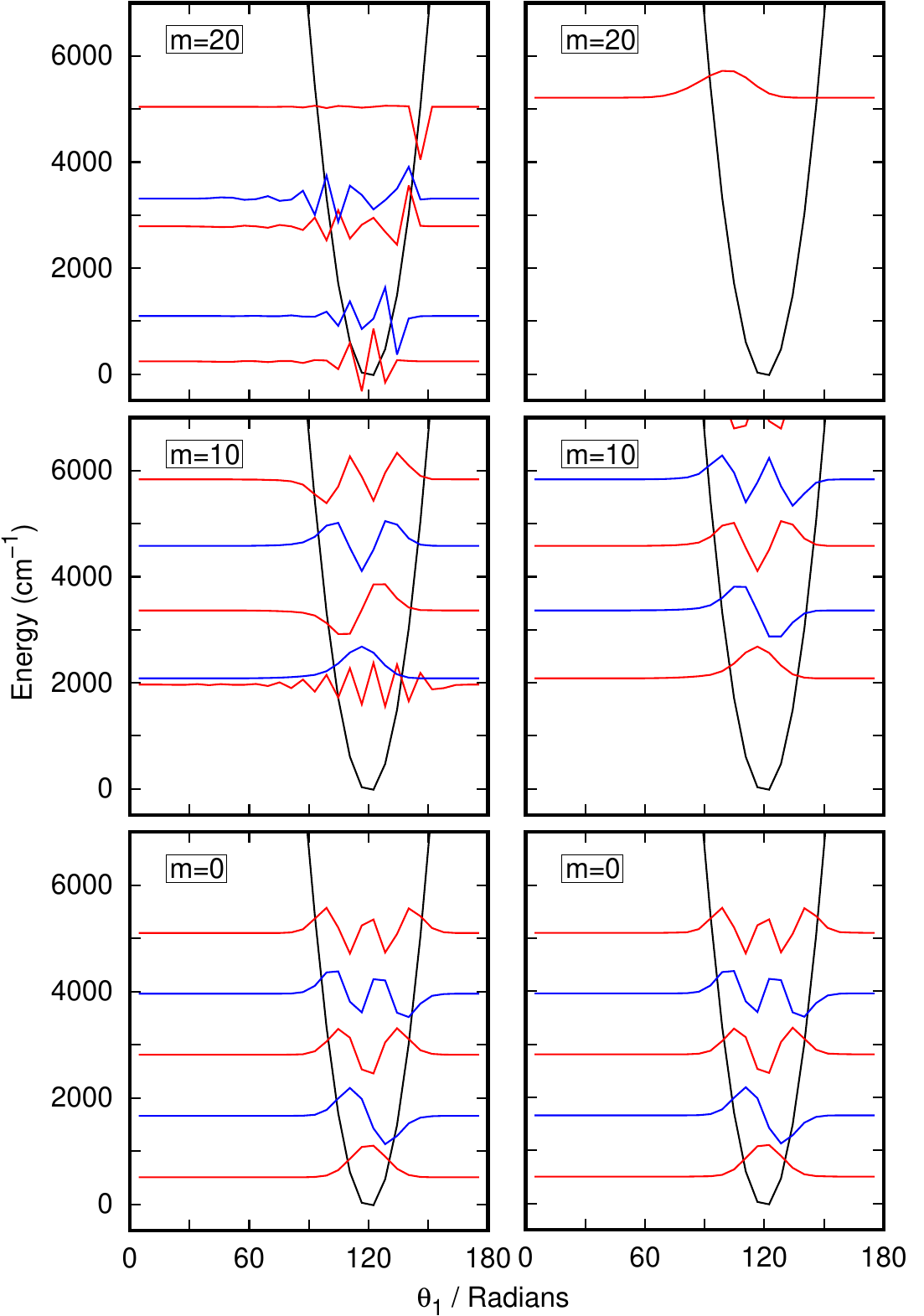}
  
  \caption{{ Monodimensional wave functions in $\theta_1$, keeping the remaining degrees of freedom at its equilibrium values, for $m$= 0 (bottom) , $m$= 10 (middle) and $m$= 20 (top) for the diagonalization in the angular grid without (left panels) and with (right panels) projection up technique. Black lines represent the potential energy, while blue/red are the angular eigen functions
  shifted to the energy of the eigen value. In this case a Gauss-Legendre angular grid of 30 points is used.
   }}    
\label{fig:spurious-up}
\end{center}
\end{figure}

To illustrate the problem and the solution of this problem
in Fig.~\ref{fig:spurious-up} the mono-dimensional eigenfunctions for $\theta_1$ are shown, which are obtained with and without
the projection technique to push up the spurious solutions for different $m$-values,
the projection of ${\bf j}_1$ in the body-fixed frame. For $m$=0, no difference is found. For $m=10$, the first eigen-function is spurious and it disappears when the projection up technique is applied. For $m$=20 the situation is even worse, and at least five spurious states appears, which are corrected and pushed up. This problem
is very notorious when calculating bound states, because the lower eigen values are mainly spurious. In wave packet propagations, this problem is in the angular representation of the Hamiltonian, and it becomes less evident, but this problem is a source of inaccuracies.


%

\end{document}